\documentclass[aps,prl,twocolumn,superscriptaddress,groupedaddress]{revtex4} 
\usepackage{graphicx}
\usepackage{dcolumn}
\usepackage{bm}
\usepackage[hidelinks]{hyperref}
\usepackage[version=3]{mhchem} 
\usepackage[T1]{fontenc}       
\usepackage{siunitx}
\usepackage{amssymb}
\usepackage{amsmath}

\begin{document}

\author{Pierre Lidon}
\affiliation{Robert Frederick Smith School of Chemical and Biomolecular Engineering, Cornell University, 120 Olin Hall, Ithaca, NY, USA}
\affiliation{CNRS, Solvay, LOF, UMR 5258, Univ. Bordeaux, 178 avenue du Dr. Schweitzer, F-33600 Pessac, France}
\email{pierre.lidon@u-bordeaux.fr}

\author{Sierra C. Marker}
\author{Justin J. Wilson}
\affiliation{Department of Chemistry and Chemical Biology, Cornell University, Baker Lab, Ithaca, NY, USA}

\author{Rebecca M. Williams}
\author{Warren R. Zipfel}
\affiliation{Department Biomedical Engineering, Cornell University, Weill Hall, Ithaca, NY, USA}

\author{Abraham D. Stroock}
\affiliation{Robert Frederick Smith School of Chemical and Biomolecular Engineering, Cornell University, 120 Olin Hall, Ithaca, NY, USA}
\affiliation{Kavli Institute at Cornell for Nanoscale Science, Physical Sciences Building, Ithaca, NY, USA}
\email{ads10@cornell.edu}

\title{Enhanced oxygen solubility in metastable water under tension}

%
%
%
%
%

\begin{abstract}
Despite its relevance in numerous natural and industrial processes, the solubility of molecular oxygen has never been directly measured in capillary condensed liquid water. In this article, we measure oxygen solubility in liquid water trapped within nanoporous samples, in metastable equilibrium with a subsaturated vapor. We show that solubility increases two-fold at moderate subsaturations ($\mathrm{RH} \sim 0.55$). This evolution with relative humidity is in good agreement with a simple thermodynamic prediction using properties of bulk water, previously verified experimentally at positive pressure. Our measurement thus verifies the validity of this macroscopic thermodynamic theory to strong confinement and large negative pressures, where significant non-idealities are expected. This effect has strong implications for important oxygen-dependent chemistries in natural and technological contexts.
\end{abstract}

\maketitle

\section{Introduction}

Many essential chemical processes in the environment and in technological applications involve reaction of solvated gases, typically in aqueous conditions: before reaching reactive sites, gas species are captured from the atmosphere by dissolution in a solvent. Examples include gases like molecular oxygen, or dioxygen, for biochemistry and respiration\cite{ingraham}, geochemistry\cite{sigel_oxygen}, corrosion\cite{gleeson_1941,mcintire_1990}, soil remediation\cite{fry_1997}, metallurgy\cite{miller_1998,greenwood_oxygen} and cathodic reductions\cite{damjanovic_1992,girishkumar_2010}; molecular nitrogen for nitrogen fixation in soils~\cite{postgate}; and carbon dioxide for photosynthesis~\cite{rabinowitch}, carbon dioxide sequestration and the formation of carbonates~\cite{berner_1983,morse_2007b}. In these scenarios, solubility defines the concentration of dissolved solute; in turn, concentration controls rates of mass transfer and reaction~\cite{levich}. In general, the solvent is present as an absorbate, in confinement within pores, or both. 

Significant research has been conducted to describe the dynamics and the equilibrium state for the adsorption of species ~\cite{bruch,thommes_2015,albasimionesco_2006,soprunyuk_2016}. The topic of gas solubility in nanoconfined solvents has more specifically received significant attention over the past two decades~\cite{mercury_2003,lassin_2005,luzar_2005,miachon_2008,peratitus_2009b,rakotovao_2010,ho_2011,clauzier_2012,soubeyrandlenoir_2012,ho_2013,ho_2015,sanchezgonzalez_2016,lassin_2016}.
Model predictions, numerical simulations as well as experiment indicate that the solubility of gases tends to be larger in nanoconfined solvents than in the bulk. Two perspectives on this oversolubility have emerged in the literature.

Starting with simulations by Luzar and Bratko~\cite{luzar_2005,bratko_2008} and experiments by Miachon et al.~\cite{miachon_2008}, a literature on oversolubility in nanoconfined solvents emerged in the context of catalysis. These studies focused on gas solubilities in synthetic mesoporous matrices wetted with organic solvents or water. These studies document dramatically increased solubilities (over ten-fold) of $\mathrm{H_2}$, $\mathrm{CO_2}$ and non-condensing organics as the confinement of the solvent increases and the solvent loading of the pore space decreases~\cite{luzar_2005,miachon_2008,ho_2011,ho_2013,ho_2015}. These investigators explain the observed oversolubility by a combination of adsorption at the pores wall and at the liquid-gas meniscus and enhanced solvation due to confinement induced layering of the solvent molecules, depending on the chemical character of the host matrix, the solvent and the solute.

Simultaneously, research led by Mercury and collaborators have emphasized the importance of capillary stress in solvents on the equilibrium constants of dissolution from both solid phases and gases in the context of geochemistry~\cite{thesis_Mercury,mercury_1997a,lassin_2005,pettenati_2008}. In the context of a solvent condensed in small pores, the capillary stress is a reduction of pressure (possibly down to negative values) controlled by Kelvin equilibrium with an unsaturated atmosphere of the solvent's vapor; concavely curved menisci pinned at the mouth of the wettable pores maintain the difference in pressure between the gas phase (positive pressure, $P_\text{g}$) and that of the solvent (reduced or even negative pressure, $P_\ell$). The reduction of pressure in the pore solvent modifies solubilities according to the rules of classical thermodynamics of bulk substances. In particular, as discussed in detail below, for solutes with positive partial molar volumes of mixing, one expects solubility to increase with decreasing solvent pressure. With a focus on water as a solvent in geological contexts (i.e. porous minerals as hosts), these investigators have used state-of-the-art descriptions ~\cite{helgeson_1981} of aqueous solutions as a function of pressure, temperature and composition to predict the dependence of solubilities on the saturation of the water vapor (relative humidity $\mathrm{RH} = P_\text{g}/P_\text{sat}(T)$ where $P_\text{sat}$ is the saturation pressure) with which the pore liquid is in equilibrium~\cite{mercury_2003}. With elegant analysis and experimentation they have shown that this Kelvin mechanism can explain measured equilibria within hydrated reactive minerals as a function of the $\mathrm{RH}$ of water vapor~\cite{lassin_2005,lassin_2016}. Of note for our study, these works provide convincing yet indirect evidence that the solubilities of $\mathrm{CO_2}$ and $\mathrm{O_2}$ increase with decreasing $\mathrm{RH}$ due to the reduced pressure within pore solvent.

Experiments to measure solubility in nanoconfined solvents have used both macroscopic and microscopic techniques. Some studies~\cite{lassin_2005,lassin_2016,ho_2011,ho_2013,ho_2015,thesis_clauzier,clauzier_2012,peratitus_2009b} used macroscopic measurements of phase equilibrium, for example by measuring the equilibrium partial pressure of $\mathrm{CO_2}$ in contact with hydrated carbonate mineral~\cite{lassin_2016} or of $\mathrm{H_2}$ in contact with a mesoporous matrix wetted by organic solvents~\cite{clauzier_2012,peratitus_2009b}. Miachon et al., on the other hand, used integrate peak size in NMR spectra of confined solutions to extract solute concentrations~\cite{miachon_2008}.

In this work, we show that the solubility of molecular oxygen ($\mathrm{O_2}$) in capillary condensed water in the pores of a wettable solid increases when the $\mathrm{RH}$ of the gas atmosphere decreases (see Fig.~\ref{fig:setup}). To do so we introduce a new microscopic technique based on phosphorescence lifetime to evaluate relative changes in oxygen concentration in liquid water in the pores of two porous media with pore radii of a few nanometers. This study represents the first direct measurement of $\mathrm{O_2}$-solubility in nanoconfinement as a function of the degree of vapor saturation. To explain our results, we adopt the Kelvin mechanism of Mercury et al., although with simpler assumptions about the properties of the solvent. This model successfully captures the observed evolutions and is consistent with previous studies on oxygen solubility in water under high positive pressure~\cite{krichevsky_1935,taylor_1978}.

We organize the remainder of our presentation as follows. First, we explain in more detail the Kelvin mechanism leading to equilibrium between pore liquid and an unsaturated vapor and its consequences on the solubility of gases. Second, we describe the experimental setup and present the phosphorescent dye and the porous media used in the study. Then, we show the results obtained from the experiments and compare them with theoretical predictions. Finally, we discuss these results relative to the aforementioned literature.


\section{Theoretical background}

Fig.~\ref{fig:setup} presents our experimental system. In the expanded view, we show an individual pore containing liquid water in equilibrium with a gas phase containing oxygen and either saturated water vapor (left) or unsaturated vapor (right). This simple representation of the pore architecture does not capture the full complexity of pores within the porous media studied here (Vycor and porous silicon) with respect to interconnectedness, distribution of pore size, or surface roughness (see section 1 in SI for more details on porous media). Nonetheless, the consideration of a simplified pore geometry (cylindrical with smooth walls ans uniform diameter) allows us to capture the relevant physics for this study. 

In the following theoretical development, we assume that the macroscopic thermodynamic properties and relations remain relevant in the pore solvent despite significant confinement (pore diameters below $\SI{10}{\nano\meter}$). By confronting the predictions of this approach with our experimental observations (see Results section), we show that, within the context of this study, these macroscopic arguments are indeed relevant. We do not exclude the possibility that nanoscale confinement has some influence on the properties of the pore liquid, we simply conclude that any such effects are not resolvable by our measurements of the variation of solubility with imposed relative humidity in pores of radius greater that $\sim \SI{1.5}{\nano\meter}$.

Classical arguments from thermodynamics predict that a capillary condensed pore liquid in equilibrium with a subsaturated vapor exists in a superheated state at reduced pressure~\cite{caupin_2012,wheeler_2009,chen_2016a} $P_\ell \ [\SI{}{\mega\pascal}]$ defined by the Kelvin-Laplace relation~\cite{wheeler_2008,gor_2017}:
\begin{equation}
P_\ell = P_\text{g} + \rho_\ell R T \ln{(\mathrm{RH})} = P_\text{g} - \frac{2\gamma \cos{\theta}}{r_\text{p}}
\label{eq:kelvin_laplace}
\end{equation}
\noindent where $P_\text{g} \ [\SI{}{\mega\pascal}]$ is the total gas pressure, $\rho_\ell \ [\SI{}{\mole\per\meter\cubed}]$ is the molar density of the liquid phase, $\mathrm{RH} = P_\mathrm{H_2O}/P_\text{sat}(T)$ is the relative humidity of the gas phase ($P_\text{sat}(T) \ [\SI{}{\mega\pascal}]$ being vapor saturation pressure), $\gamma \ [\SI{}{\joule\per\meter\squared}]$ is the surface tension (for pure bulk water, $\gamma=\SI{72}{\milli\joule\per\meter\squared}$ but this value could depend on the curvature of the meniscus and of the concentration of dissolved gases) of the liquid and $\theta \ [\SI{}{\radian}]$ is the contact angle with the solid at the pore mouth. Equation~\eqref{eq:kelvin_laplace} predicts that the pressure $P_\ell$ in the pore liquid decreases (and eventually becomes negative) as $\mathrm{RH}$ drops; this state persists until cavitation occurs or until the contact angle of the meniscus reaches its receding value. Modest changes in $\mathrm{RH}$ should induce large changes in the pressure experienced in the adsorbate: for instance, while pressure in saturated conditions ($\mathrm{RH} = 1$) is about $P_\ell = \SI{0.1}{\mega\pascal}$, it decreases to $P_\ell = -\SI{80}{\mega\pascal}$ for a moderate unsaturation $\mathrm{RH}=0.5$.

We~\cite{vincent_2014,vincent_2016,vincent_2017a} and other groups~\cite{gruener_2009,huber_2015} have recently shown that predictions of Eq.~\eqref{eq:kelvin_laplace} hold to extremes of confinement (pore radius down to $\SI{1.4}{\nano\meter}$) and pressure ($P_\ell = -\SI{70}{\mega\pascal}$). In the following, we use Eq.~\eqref{eq:kelvin_laplace} to interpret responses to changes in $\mathrm{RH}$  in terms of changes in solvent pressure. This approach provides a simple route to pursue an unprecedented study of solubility at reduced pressures, extending previous studies conducted under elevated pressures~\cite{taylor_1978}. 

The effect of solvent pressure on the solubility can be modelled by using simple arguments based on macroscopic thermodynamics. If we assume that the partial molar volume of oxygen in water $v_2 \ [\SI{}{\meter\cubed\per\mole}]$ is independent of pressure, Gibbs-Duhem equation can be easily integrated to give the change of chemical potential with pressure. If we further assume that the gas phase is ideal, we get the Krichevsky and Kasarnovsky equation~\cite{krichevsky_1935}:
\begin{equation}
\frac{c(P_\ell)}{c_\text{sat}} = \frac{\rho_\ell(P_\ell)}{\rho_{\ell,\text{sat}}} \mathrm{e}^{-v_2 (P_\ell - P_\text{sat}) / RT}.
\label{eq:KK_equation}
\end{equation}
\noindent We provide a derivation of this relation in section 4 of SI. These arguments assume that the pore liquid can be treated as having a well-defined pressure and bulk behavior despite the significant confinement imposed by the walls of the nanoscale pores. For the pore dimensions used in this study ($r_\text{p} = \SI{1.4}{\nano\meter}$ and $\SI{3.8}{\nano\meter}$), our previous work ~\cite{vincent_2016} and that of others ~\cite{bocquet_2010} suggest that this assumption of macroscopic thermodynamic behavior should hold. The success of this treatment in describing our data (see Results section) serves as a test of its consistency. In section 4 of SI, we also provide a more general relation directly between the imposed $\mathrm{RH}$ and the solubility. Variations of molar volume ($v_2$) with pressure can be accounted for by using van der Waals-like equations of state~\cite{widom_2012,cerdeirina_2016}.


\section{Materials and methods}

\begin{center}
\begin{figure}[!htb]
\centering
\includegraphics{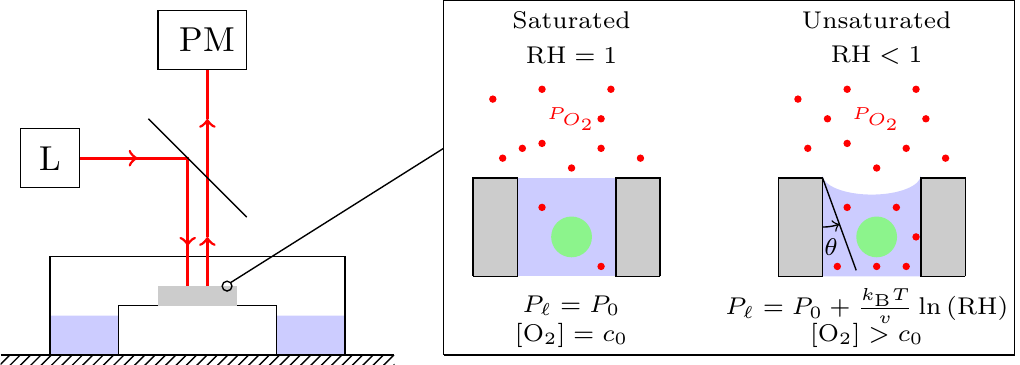}
\caption{Measuring relative concentrations of oxygen in pore liquid under tension with phosphorescence lifetime. A porous sample wetted with an aqueous solution of phosphorescent dye (green circles in the expanded view) was placed in a box of controlled atmosphere of water vapor, oxygen (red circles in the expanded view) and other inert gases, in equilibrium with osmotic solutions imposing relative humidity $\mathrm{RH}$. Phosphorescence lifetime was accessed by illuminating the sample with pulsed laser (L) and measuring the decay of phosphorescence signal with a photomultiplier (PM).\label{fig:setup}}
\end{figure}
\end{center}

The experimental setup is described in Fig.~\ref{fig:setup}. Initially, we filled a dry porous sample with a solution of oxygen-sensitive phosphorescent dye with concentration~$\sim \SI{100}{\micro\mole\per\liter}$: we placed a drop of solution on the sample and left it at rest for about~$\SI{15}{\minute}$ to allow filling by capillary imbibition. After carefully drying the surface of the sample, we placed it in a closed chamber with controlled humidity $\mathrm{RH}$ and oxygen pressure $P_\mathrm{O_2}$, and allowed equilibration of pore liquid with the atmosphere for~$\sim \SI{5}{\minute}$ before measurements. We used two different porous solids -- porous glass (pore radius $r_\text{p}=\SI{3.8(3)}{\nano\meter}$) and porous silicon ($r_\text{p}=\SI{1.4(1)}{\nano\meter}$) -- to host the condensed water. Some details on the porous samples are given in section 1 in SI.

The composition of the atmosphere for each experiment was either air or a mixture of oxygen and nitrogen at atmospheric pressure $P_\text{g}=\SI{1}{\bar}$ and with the ratio controlled by a gas mixer (Oxydial); the mixer was calibrated with respect to the Clark electrode. The humidity of the atmosphere was imposed by the saturation vapor pressure of salt solutions placed in the box; no direct contact occurred between this solution and the porous sample. In this experiment, solutions of sodium or lithium chloride at various concentrations were used to impose liquid pressures ranging from~$P_\ell = P_\text{sat}$ (pure water, saturated atmosphere) to~$\SI{-70}{\mega\pascal}$. For each solution and before every experiment, water potential was measured using a chilled mirror hygrometer (WP4C, Meter Group~\cite{aqualab}). The experiment was performed at a temperature of $T=\SI{20}{\celsius}$: in the following, all numerical results and theoretical predictions will be given at his temperature.


\onecolumngrid
\twocolumngrid

\begin{figure*}[!htb]
\centering
\includegraphics{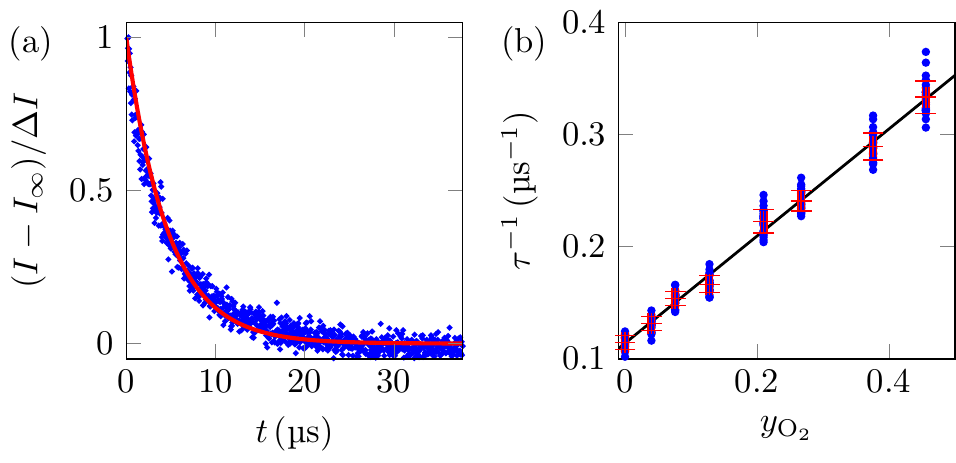}
\caption{Phosphorescence lifetime of dye in pore liquid. (a) Example of phosphorescence signal and exponential fit, with lifetime~$\tau=\SI{4.7}{\micro\second}$, amplitude $\Delta I=479$ and noise level $I_\infty = 135$. (b) Evolution of lifetime in Vycor with mole fraction of oxygen, $y_\mathrm{O_2}$, in saturated atmosphere. Blue dots are the measured lifetimes at different positions on the porous glass ($N=25$), red crosses represent the average lifetime and errorbars correspond to standard deviations over the sample. Black line corresponds to a linear fit~\eqref{eq:stern_volmer_gas} with~$\tau_0=\SI{8.7(3)}{\micro\second}$ and~$k_\text{sat}=\SI{0.48(2)}{\per\micro\second}$.   \label{fig:optical_method}}
\end{figure*}

To measure the concentration, $c \ [\SI{}{\mole\per\meter\cubed}]$, of dissolved oxygen in pore liquid water, we exploited the oxygen quenching of the phosphorescence of specific dyes~\cite{esipova_2011,choi_2012}: we used a rhenium based phosphor~\cite{marker_2018} synthesized by our team (details in section 3 of SI). This dye presents the advantage over commercial ones by yielding significant luminescence signal, and being water soluble and small enough to fit within the nanoscale pores. By fitting the luminescence signal following pulsed laser excitation of the phosphor, we measured the phosphorescence lifetime, $\tau \ [\SI{}{\second}]$ of the dye (see Fig.~\ref{fig:optical_method}(a) and section 2 in SI for more details on the optical setup). 

For both substrates, we observed variations of lifetime when we performed measurements at different locations across the samples. However, the size of the laser spot (diameter $\sim \SI{200}{\micro\meter}$) was larger than the typical pore size and we thus expected to observe an homogeneous value. It is known that lifetime of fluorescent dyes can display fluctuations due to molecular conformation~\cite{tinnefeld_2000,esipova_2011}, but this effect should not affect significantly our measurement as the spot size is larger than the typical pore size. However, it has been observed that Stern-Volmer parameters can display significant variations on micrometric scales on a priori homogeneous surfaces~\cite{bedlekanslow_2000,kneas_2000}. Furthermore, as was seen with other dyes~\cite{kennelly_1985,shi_1985}, we observed a significant variation of lifetime between the different porous media and with respect to bulk solution (see section 3 in SI for more details): this observation hints at a strong interaction between the dye and the solid glass matrix which could explain the observed spatial variations, as small variations of the glass properties on large scale could significantly impact measured lifetimes.  In order to moderate the impact of these fluctuations, we repeated measurements for each value of $\mathrm{RH}$ at different places of the samples ($N=25$) and computed average lifetimes. Uncertainties are represented by the standard deviation of the measured series. 

We expect the relationship between the lifetime, $\tau$, and the concentration, $c$, of oxygen in the pore liquid to be given by the Stern-Volmer relation:
\begin{equation}
\frac{1}{\tau} = \frac{1}{\tau_0} + K c.
\label{eq:stern_volmer_liquid}
\end{equation}
\noindent where $\tau_0 \ [\SI{}{\second}]$ and $K \ [\SI{}{\meter\cubed\per\mole\per\second}]$ are parameters to be determined by calibration which depend on the nature of the substrate. The concentration is related to the oxygen pressure through Henry's law $c = \mathcal{H} P_\mathrm{O_2}$, so we predict:
\begin{equation}
\frac{1}{\tau} = \frac{1}{\tau_0} + k y_\mathrm{O_2}
\label{eq:stern_volmer_gas}
\end{equation}
\noindent where $k=\mathcal{H} K P_\text{g}$ and $y_\mathrm{O_2}=P_\mathrm{O_2}/P_\text{g}$ is the mole fraction of $\mathrm{O_2}$ in the atmosphere.

To calibrate the sensor, we filled the experimental cell with a mixture of oxygen and nitrogen with a controlled ratio and at atmospheric total pressure $P_\text{g}=P_\text{atm}$, saturated with water vapor ($\mathrm{RH}=1$). The lifetimes we obtained for different mole fractions of oxygen in the gas phase is presented in Fig.\ref{fig:optical_method}(b). For both substrates, the equation~\eqref{eq:stern_volmer_gas} describes satisfactorily our measurements and this calibration allows us to determine the values of $\tau_0$ and $k_\text{sat}=\mathcal{H}_\text{sat} K$, where subscript $\text{sat}$ indicates that this value is valid at saturation. Values of the parameters for the two substrates are given in section 3 in SI.

For any value of $\mathrm{RH}$, we have
\begin{equation}
\frac{1}{\tau} = \frac{1}{\tau_0} + k_\text{sat} y_\mathrm{O_2} \frac{c}{c_\text{sat}}.
\label{eq:stern_volmer_final}
\end{equation}
\noindent As we control $y_\mathrm{O_2}$ and by assuming that the determined values of $\tau_0$ and $k_\text{sat}$ do not depend on the pressure in the liquid, our measurements of lifetime give us access to $c/c_\text{sat}$. In other words, we can track relative changes in the concentration of $\mathrm{O_2}$ in the pore liquid with this technique; we cannot determine absolute concentrations.

\section{Results}

Fig.~\ref{fig:conc_vs_RH} represents the observed evolution of the relative concentration of dissolved oxygen in the pore liquid (aqueous solution of phosphor) for fixed mole fraction in the gas phase ($y_\mathrm{O_2}=0.21$) and various values of $\mathrm{RH}$. We used calibration curves as in Fig.~\ref{fig:optical_method} to evaluate the relative concentration from phosphorescence lifetimes, with the value at saturation ($\mathrm{RH}=1$), $c_\text{sat}$, as a reference. We report the imposed $\mathrm{RH}$ on the top axis and the corresponding pressure in the pore liquid (calculated with Eq.~\eqref{eq:kelvin_laplace}) on the bottom axis.

\begin{figure}[!htb]
\centering
\includegraphics{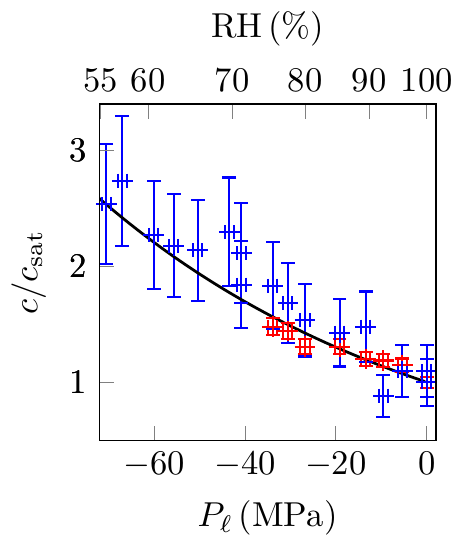}
\caption{Evolution of oxygen concentration $c$ with liquid pressure $P_\ell$ in metastable water, with respect to its value $c_\text{sat}$ at saturation. Red symbols correspond to measurements in Vycor glass and blue symbols to measurements in porous silicon. Black line is the prediction of the simple model (Equation~\eqref{eq:KK_equation}). \label{fig:conc_vs_RH}}
\end{figure}

The range of $\mathrm{RH}$ (or pore pressure) over which we report concentrations for each sample type ($\sim 0.75 \leq \mathrm{RH} \leq 1$ for Vycor and $\sim 0.55 \leq \mathrm{RH} \leq 1$ for porous silicon) were limited from below by the disappearance of detectable phosphorescence for drier conditions. We interpret this loss of signal as occurring at the point at which the pore liquid receded into the pores when the stress implied by the Kelvin equation ($\rho_\ell R T \ln{\mathrm{RH}}$ in Eq.~\eqref{eq:kelvin_laplace}) exceeded the maximum capillary stress sustainable in the pores~\cite{rasmussen_2010} ($2\gamma \cos{\theta_\text{r}} / r_\text{p}$ in Eq.~\eqref{eq:kelvin_laplace} where $\theta_\text{r}$ is the receding contact angle of water on the pore wall). For Vycor, with $r_\text{p}=\SI{3.8}{\nano\meter}$, we expect this drying transition to occur at $\mathrm{RH}=0.8$; for porous silicon, with $r_\text{p}=\SI{1.4}{\nano\meter}$, we expect it at $\mathrm{RH}=0.54$. In both cases, we use a reasonable value $\theta_\text{r}=\SI{40}{\degree}$ and find reasonable agreement with the observed transitions.

We now remark the good agreement in the measured values of relative concentration as a function of $\mathrm{RH}$  for the two substrates over the shared portion of their ranges ($\sim 0.75 \leq \mathrm{RH} \leq 1$). We further note the large (over two-fold) increase in relative concentration of dissolved $\mathrm{O_2}$ as the vapor goes from saturation to $\mathrm{RH} = 0.55$. We conclude that a modest degree of subsaturation in the vapor phase, commonly encountered in our environment, has a dramatic effect on the solubility of $\mathrm{RH}$ of $\mathrm{O_2}$ in the pore liquid.

The black line in Fig.~\ref{fig:conc_vs_RH} represents the prediction of Krichevsky and Kasarnovsky (Equation~\eqref{eq:KK_equation}) for a literature value for the molar volume~\cite{taylor_1978,zhou_2001} of $\mathrm{O_2}$ in water: $v_2=\SI{3.21e-5}{\meter\cubed\per\mole}$. Predictions of the more sophisticated model detailed in section 5 of SI match those of the simple model within a few percent, well within the experimental uncertainty. Thus, our theory agrees quantitatively with experimental results. Importantly, we employed no fitting parameter in performing this comparison. Interestingly, the same Krichevsky and Kasarnovsky equation had previously been used to describe the evolution of oxygen solubility at positive pressures~\cite{taylor_1978}; our experiment thus extends its range of validity by a factor two and, for he first time, into the regime of negative pressure. 

\section{Discussion}

\subsection{Comparison with literature}

Given the structural and chemical distinctions between the two substrates, the agreement between data acquired for both of them and our model for a wide range of vapor saturations demonstrates:
\begin{enumerate}
\item the appropriateness of interpreting the impact of subsaturation as the impact of the thermodynamic state of the pure fluid imposed by metastable equilibrium with its vapor (Equation~\eqref{eq:kelvin_laplace}) and not of the detailed characteristics of the host matrix, and
\item the relevance of bulk thermodynamics relations for multi-phase, multi-component equilibrium for nano-confined pore liquids.
\end{enumerate}

This success supports the perspective put forward by Mercury, Lassin et al.~\cite{mercury_2003,mercury_2004,lassin_2005,lassin_2013,lassin_2016} that the Kelvin stress within the pore fluid plays an essential role in defining the solubility of gases in unsaturated porous media. Our model relies on a far simpler description of the thermodynamic properties of the solute and solvent than that employed by these authors~\cite{helgeson_1981,lassin_2016}. However, the matrices used in our study should be completely unreactive (except for the possible deprotonation of silanol groups) such that pore liquid remains essentially pure water and ion effects should be minimal. In contrast, within the geological samples of interest to them, account for these effects was necessary. The good agreement between Krichevsky and Kasarnovsky equation and a more complete model with an equation of state for the mixture of oxygen and liquid water (see section 5 in SI) can be explained by the small variation of the partial molar volume $v_2$ of oxygen with pressure~\cite{mercury_2003}.


We further consider the interplay between Kelvin mechanism and the studies following experiments by Miachon~\cite{miachon_2008} and Ho~\cite{ho_2011} in which the effects of the saturation of the solvent vapor were not explored. Our experiments are restricted to a situation in which the porous solid remains completely filled with the liquid and the liquid achieves equilibrium with a nanoscopic deformation of the meniscii pinned at the pore mouths (see expanded view in Fig.~\ref{fig:setup}). In this filled state, the extent of liquid-solid and liquid-vapor interfaces within the host stays constant as we vary $\mathrm{RH}$, such that we expect the interfacial effects involved by Miachon and Ho should not impact the variation of relative solubility, $c/c_\text{sat}$, that we report here. On the other hand, we expect that this mechanism should impact the absolute concentrations in our host media (e.g. the concentration at vapor saturation, $c_\text{sat}$, may be larger than in bulk water). However, due to our calibration procedure, we only measure directly the ratio $c/c_\text{sat}$ of concentration at a given humidity over its reference saturated value. We do observe a significant discrepancy between the average lifetime of the dye in bulk solution and in pores under saturated conditions, however, we cannot interpret this observation as an effect of confinement on solubility as the behaviour of the dye itself is likely affected by the confinement as we noted previously (see section 3 in SI for a comparison of lifetimes in solid substrates and in bulk solutions).

We cannot exclude that the confinement induced oversolubility mechanism would play a role in our system at low saturations for which gas and liquid would be interspersed through the whole pore space. Our method of observation was not sensitive enough to obtain phosphorescence signal in this regime. While the mechanism proposed by Miachon and Ho is likely to be dominant in the very dry regime, where the liquid only remains as an adsorbate on the pores walls, it should be corrected by the Kelvin mechanism in the more humid regime when capillary condensation starts to occur.

\subsection{Absence of anomalous hydrophobic effect}

Under the relative humidities involved in our study, we predict that liquid pressure reaches $\sim -\SI{70}{\mega\pascal}$, for which density should decrease by about $\sim \SI{4}{\percent}$ with respect to saturation.  It is surprising that our simple model, based on thermodynamic assumptions of ideality and bulk properties of water, holds in this notably stretched state and at such confinement, in pores of diameter of the order of a few molecular diameters.

In particular, the solubility of species in water under tension has been investigated numerically in the context of hydrophobic interactions, involved for instance in protein folding~\cite{chen_2017b}. In this regime, there are strong numerical clues that the properties of solvent should be significantly altered and that an anomalous solvation effect occurs for hydrophibic molecules~\cite{rajamani_2005}. 

Oxygen has not been specifically studied with respect to the hydrophobic effect but its apolar nature and the non monotonous behavior of its solubility with temperature~\cite{battino_1966,clever_2014} hint at a hydrophobic character. The agreement with our theory is thus particularly striking. Further studies involving more hydrophobic and larger molecules would be of strong interest. 

\subsection{Impact on physico-chemical processes}

Such a dependency of the solubility of oxygen on $\mathrm{RH}$ for capillary condensed water is expected to impact various physico-chemical processes. In natural conditions, capillary condensed water is mostly found in soils: there, dissolved oxygen can be beneficial, for instance in agriculture as it enhances root growth~\cite{geisler_1967,nobel_1989} and microbial activity~\cite{molz_1986}, or detrimental, for instance as it triggers the formation of polluting sulphidic compounds in mine tailings~\cite{yanful_1993}. Significant work has been done on the modelling of oxygen transport within unsaturated soils~\cite{chesnaux_2009} and incorporation of our result could help to reach more quantitative predictions.

Electrochemical transformations represent important class of processes susceptible to be impacted by this phenomennon. Dry corrosion of metals, which happens in contact with an unsaturated medium, is one of the most widely studied physico-chemical phenomenon due to the enormous industrial cost it represents for the maintenance of pipes and structural materials~\cite{corrosion_report}. Many studies have reported effects of the relative humidity in the gas phase on corrosion rates in several contexts like atmospheric corrosion ~\cite{vernon_1931,mansfeld_1976,stratmann_1990a,stratmann_1990b,tsuru_1995}, corrosion by hot steam~\cite{deal_1963} or soil corrosion~\cite{gupta_1979,gardiner_2002}. In all these cases, the metals are covered by a thin liquid film and any asperity of the surface may host a pocket of water under tension. Our findings are thus highly relevant in this context and may be a key phenomenon for the understanding of these observations.


In the context of research for sustainable power sources, a significant effort is devoted to fuel cells where oxygen is delivered in a gaseous phase to serve as an oxidant. Among these various technologies, air breathing fuel cells operate with reactants supplied to the electrodes by gas flows with controlled $\mathrm{RH}$ to maintain hydration of the electrodes~\cite{ferrigno_2002,coz_2016}. At high current density, the efficiency is limited by oxygen transport~\cite{kjeang_2009,jayashree_2005,owejan_2014,weber_2014} and it has been shown that the relative humidity of the gas has an impact on this efficiency~\cite{chu_1999,jayashree_2005,simon_2017}. Our results offer a possible basis for rationalizing this observation. Moreover, it opens the way for new design of porous cathodes that could exploit this enhanced oxygen solubility in unsaturated conditions to improve significantly the performance of this technology.

\section{Conclusions}

We have shown in the present work that the solvation of oxygen in adsorbed water is strongly influenced by the relative humidity of the surrounding atmosphere and that this effect can be described quantitatively by thermodynamic models using bulk properties of water. This success extends the conclusions made by our team~\cite{vincent_2014,vincent_2016,vincent_2017a} and others~\cite{gruener_2009,huber_2015} that the  bulk thermodynamic behaviors of liquid water persist to strong confinement and large tension. The Krichevsky and Kasarnosky theory is not restricted to dioxygen and predicts a similar exponential variation of the solubility of any gas with pressure. Consistent with Le Chateliers's principle, the solubility of solutes with positive partial molar volumes of mixing should increase with increasing tension while the solubility of solutes with negative partial molar volumes of mixing should decrease. Beyond this dependence on the partial molar volume, this prediction does not depend on the identity or microscopic details of the solvent and solute. Further work should confront this prediction by considering other molecules that display a strong hydrophobic character and for which we may expect qualitative changes of the solvent properties of pore-confined liquid water under tension.

This modification of gas solubility in adsorbed water is likely to affect important chemical processes like corrosion and it may be a missing ingredient for modeling in many situations in which dependence on $\mathrm{RH}$ has been observed. More interestingly, this study calls for the development of chemical processes exploiting capillary condensed water as a tunable solvent in the context of green chemistry: while most of studies on the effect of pressure have focused on the supercritical regime, our study shows that pressure quantitatively affects the solubility in water in a regime easier to access in practice and opens the path to the study of chemistry in metastable water, interesting for both fundamental and applied perspectives. 

\textit{Ackowledgements -}Experiments presented in this article have been conducted in the Biotechnology Resource Center Imaging Facility at Cornell University. The authors thank Prof. Benjamin Widom for interesting discussion on the theory, Hanwen Lu for discussions and characterization of Vycor samples and Olivier Vincent for making the porous silicon sample. The authors acknowledge an anonymous reviewer for the suggesting the relation between solubility and relative humidity (included in section 4 of SI). The authors acknowledge funding from Air Force Office for Scientific Research (FA9550-15-1-0052) and Cornell College of Engineering. Substrates were prepared, in part, in the Cornell Nanoscale Facility, an NNCI member supported by NSF grant ECCS-1542081.


\providecommand{\latin}[1]{#1}
\makeatletter
\providecommand{\doi}
  {\begingroup\let\do\@makeother\dospecials
  \catcode`\{=1 \catcode`\}=2 \doi@aux}
\providecommand{\doi@aux}[1]{\endgroup\texttt{#1}}
\makeatother
\providecommand*\mcitethebibliography{\thebibliography}
\csname @ifundefined\endcsname{endmcitethebibliography}
  {\let\endmcitethebibliography\endthebibliography}{}

\newpage

\section{Supporting Information}

\section*{1. Porous materials}

Two different porous materials have been used in the experiment:
\begin{itemize}
\item circular disk (radius~$\sim \SI{2}{\milli\meter}$, thickness $\sim \SI{500}{\micro\meter}$) of Vycor 7930 porous glass. Before every experiment, the sample was cleaned by immersion for $\sim \SI{15}{\minute}$ in boiling hydrogen peroxide ($\SI{30}{\percent}$ by mole in water). Pore size distribution was determined by nitrogen adsorption porosimetry. It displays a sharp maximum at radius~$\SI{3.8}{\nano\meter}$ with a half width of~$\SI{0.3}{\nano\meter}$.
\item a thin layer ($\sim \SI{5}{\micro\meter}$-thick) of nanoporous silicon. This layer was produced by anodization in a similar manner to that previously described by us~\cite{vincent_2014} but with thermal oxidation at higher temperature ($\SI{2}{\minute}$ at~$\SI{800}{\celsius}$ followed by~$\SI{30}{\second}$ at~$\SI{900}{\celsius}$ in pure oxygen) after synthesis. The pore size distribution of the sample used in this study was not measured but we expect it to be similar to that measured previously~\cite{vincent_2014}: average radius~$\SI{1.4}{\nano\meter}$ and half width $\SI{0.3}{\nano\meter}$, with possibly slightly smaller average size due to the stronger oxidation.
\end{itemize} 

For both systems, there is a three dimensional disordered and interconnected pore structure. However, due to scattering of the laser beam, we only probe liquid in pores at or close to the surface of the sample in our experiment and there is thus no impact of this complex organization on our results.
 
\section*{2. Optical setup}

Laser excitation for the phosphorescence lifetime measurements was provided by pulsing the $\SI{405}{\nano\meter}$ laser line from a four-line iChrome MLE laser (Toptica Photonics AG, Munich, Germany). The diode laser in the iChrome was triggered by a DG535 Digital Delay/Pulse Generator (Stanford Research, Sunnyvale, CA) at $\SI{100}{\kilo\hertz}$ and delivered $\SI{100}{\nano\second}$ FWHM $\SI{770}{\nano\meter}$ excitation pulses. The $\SI{405}{\nano\meter}$ pulses were fiber-delivered to a modified BioRad MRC-600 confocal scanhead mounted on an Olympus BX-51 upright microscope stand. Excitation pulses were delivered through an Olympus 4x/0.28 NA (XLFLUOR4X/340) objective lens focused on the sample and the phosphorescence collected back though the confocal scanhead. The phosphorescence was routed to one of the MRC-600's internal multi-alkali  photomultiplier tubes (PMT) after passing through a $\SI{470}{\nano\meter}$ long pass filter (HQ470lp, Chroma Technology, Bellows Falls, VT) and the confocal pinhole which was set to $\sim 1$ Airy disk (optical sectioning was $\SI{9.0}{\micro\meter}$).  Photon counts from the PMT output were collected in $\SI{40}{\nano\second}$ time bins using a SR430 Multi-channel scaler (Stanford Research, Sunnyvale, CA). Data was transferred to a PC via the SR430 GPIB bus. As displayed in Fig.2(a) in main text, lifetime $\tau$ is obtained by fitting the luminescence intensity with a single exponential decay
\begin{equation}
I(t) = \Delta I \mathrm{e}^{-t/\tau} + I_\infty.
\label{eq:exponential_decay}
\end{equation}

\section*{3. Luminescent dye}

\begin{center}
\begin{figure}[h]
\includegraphics[height=4cm]{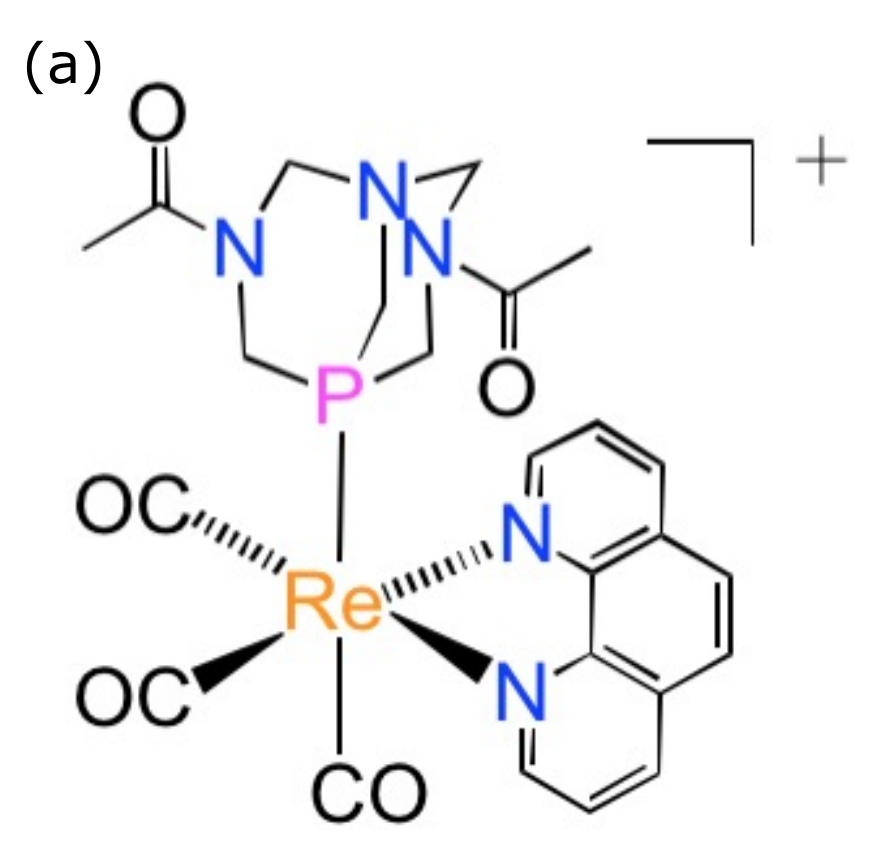}
\includegraphics[height=4cm]{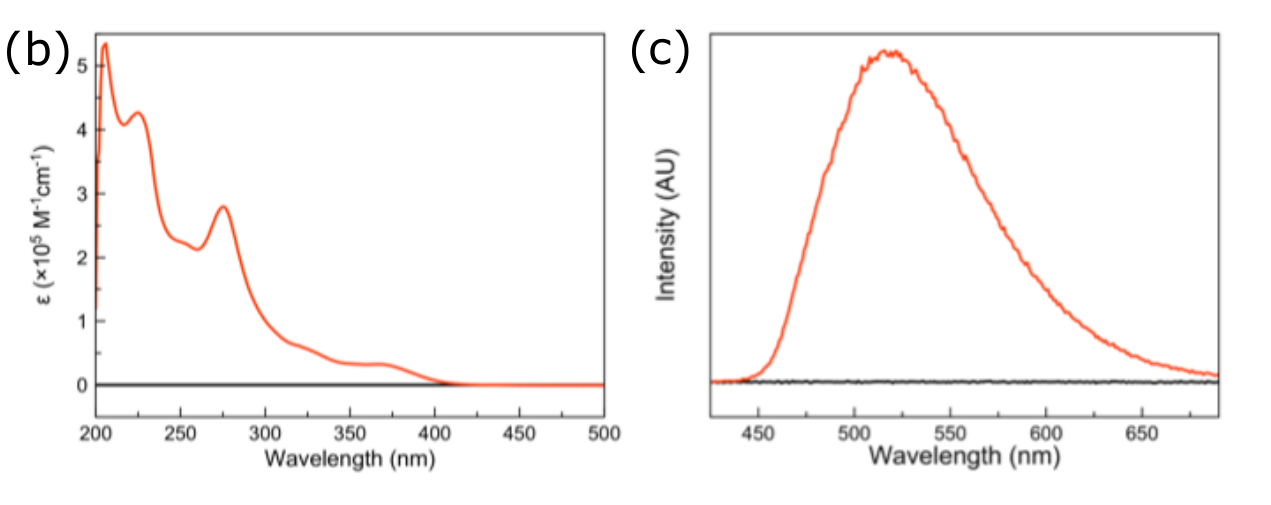}
\caption{(a) Structure of the phosphorescent dye used as an oxygen probe. (b) UV-visible absorption spectrum of the dye. (c) Luminescence emission of the dye for an excitation at $\lambda=\SI{350}{\nano\meter}$. \label{fig:rhenium_dye}}
\end{figure}
\end{center}

The dye used in this experiment is a water soluble rhenium-based complex whose structure is described in Fig.~\ref{fig:rhenium_dye}(a). Details about its synthesis and characterization are given in Marker et al.~\cite{marker_2018}

\onecolumngrid
\twocolumngrid

Analytical photochemical measurements were performed using a Newport Mercury/Xenon Arc Lamp. The light output was modulated using a combination of a Newport heat absorbing glass filter ($50.8 \times \SI{50.8}{\milli\meter}$) with infrared cut-off (Schott KG5 filter glass), a Newport mercury line bandpass filter ($\SI{25.4}{\milli\meter}$, center wavelength $\SI{365(2)}{\nano\meter}$), and a Newport visible absorbing filter ($50.8 \times \SI{50.8}{\milli\meter}$, center wavelength $\SI{340}{\nano\meter}$) made of dark optical glass to isolate monochromatic $\SI{365}{\nano\meter}$ light. The UV-visible absorption spectrum of the dye and its emission spectrum for excitation at $\lambda=\SI{350}{\nano\meter}$ are respectively displayed on Fig.~\ref{fig:rhenium_dye}(b) and (c).

As explained in main text, the luminescence lifetime of the dye is related to the concentration of oxygen in pore liquid via the equation
\begin{equation}
\frac{1}{\tau} = \frac{1}{\tau_0} + k y_\mathrm{O_2}.
\label{eq:stern_volmer_gas_2}
\end{equation}
\noindent The calibration procedure to determine the parameters $\tau_0$ and $k$ has been described in main text. In Vycor, we obtained $\tau_0=\SI{8.7(3)}{\micro\second}$ and $k_\text{sat}=\SI{0.48(2)}{\per\micro\second}$ while in porous silicon, we obtained  $\tau_0=\SI{1.8(1)}{\micro\second}$ and $k_\text{sat}=\SI{0.6(1)}{\per\micro\second}$.

As evoked in the main text, we suspect that interactions between the dye and the wall modify significantly $k_\text{sat}$ and $\tau_0$. This explains the discrepancies of the values obtained for these parameters in the two porous media and justifies the necessity to perform in situ calibrations. 

This is further supported by a few complimentary tests. In bulk water under ambient conditions ($y_\mathrm{O_2}=0.21$), we measured a lifetime $\tau=\SI{2.3(1)}{\micro\second}$, to be compared with  $\tau=\SI{4.5(2)}{\micro\second}$ in Vycor and $\tau=\SI{1.5(1)}{\micro\second}$ in porous silicon in the same conditions. We thus observe an notable discrepancy between lifetime measured in bulk solution and under confinement. For a phosphate buffered solution ($\mathrm{pH}=7.4$, concentration $\SI{1}{\milli\mole\per\liter}$) we measured $\tau=\SI{1.9}{\micro\second}$ in ambient conditions and $\tau_0=\SI{5.6}{\micro\second}$ in a solution saturated with nitrogen.

\section*{4. Derivation of Krichevsky and Kasarnovsky equation}

Let us consider the equilibrium between an ideal solution of molecular oxygen in water, of concentration $c$ and pressure $P_\ell$, and a gas phase, of total pressure $P_\text{g}$ with a mole fraction $y_\mathrm{O_2}$ in oxygen and a relative humidity, $\mathrm{RH}$. We want to compare the concentrations $c$ and $c_\text{sat}$ of oxygen between the considered situation and the saturation where $\mathrm{RH}=1$ and $P_\ell = P_\text{sat}$ (see Eq. (1) in main text). The temperature $T$ is kept constant.

In both situations, the mole fraction of oxygen and the total pressure in the gas phase are the same, so if we assume the vapor is ideal, the chemical potential of oxygen in the gas is the same in the two situations: at equilibrium, this implies that the chemical potential of dissolved oxygen in the liquid phase is the same too, i.e.:
\begin{equation}
\mu_{2,\ell} (P_\ell,T,c) = \mu_{2,\ell} (P_\text{sat},T,c_\text{sat}).
\label{eq:equilibrium_condition}
\end{equation}

First, we can expand the term on the left hand side with respect to pressure:
\begin{equation}
\mu_{2,\ell} (P_\ell,T,c) = \mu_{2,\ell} (P_\text{sat},T,c) + \int_{P_\text{sat}}^{P_\ell} \left(\frac{\partial \mu_{2,\ell}}{\partial P}\right)_{T,c_\text{sat}} \mathrm{d} P.
\end{equation}
\noindent If we introduce the partial molar volume of oxygen $v_\mathrm{O_2} =\partial_P \mu_{2,\ell}$ and assume it is independent of pressure, we obtain
\begin{equation}
\mu_{2,\ell} (P_\ell,T,c) = \mu_{2,\ell} (P_\text{sat},T,c) + v_2(P_\ell - P_\text{sat}).
\end{equation}

If we assume the solution is ideally dilute, we can write
\begin{equation}
\left\{ \begin{array}{l}
\mu_{2,\ell} (P_\text{sat},T,c) = \mu^\infty_{2,\ell} (P_\text{sat},T) + RT\ln{(c/\rho_\ell)} \\
\mu_{2,\ell} (P_\text{sat},T,c_\text{sat}) = \mu^\infty_{2,\ell} (P_\text{sat},T) + RT\ln{(c_\text{sat}/\rho_\text{sat})}
\end{array} \right.
\end{equation}
\noindent where $\mu^\infty_{2,\ell}(P_\text{sat},T)$ is the chemical potential of molecular oxygen in the hypothetical infinitely dilute solution, which is independent of the concentration of oxygen, and $\rho_\ell$ is the molar density of the liquid, which depends on its pressure.

By combining these expressions, we obtain the Krichevsky and Kasarnovsky equation:
\begin{equation}
\frac{c(P_\ell)}{c_\text{sat}} = \frac{\rho_\ell(P_\ell)}{\rho_{\ell,\text{sat}}} \mathrm{e}^{-v_2 (P_\ell - P_\text{sat}) / RT}.
\end{equation}

This formula has been initially derived in a context of elevated pressure (i.e. $P_\ell > P_\text{sat}$), with bulk liquid. In our situation, it could be questionable due to confinement: because of solid/liquid interaction, the pressure becomes heterogeneous and anisotropic at nanoscale and is thus ill-defined. It could be better not to refer to pressure but rather to only consider chemical potential and relative humidity which does no suffer from wall effect. The equation (\ref{eq:equilibrium_condition}) then becomes
\begin{equation}
\mu_{2,\ell} (\mathrm{RH},T,c) = \mu_{2,\ell} (\mathrm{RH}=1,T,c_\text{sat}).
\end{equation}
\noindent The most general relation that we can get between $\mathrm{RH}$ and the solubility is
\begin{equation}
\frac{c(\mathrm{RH})}{c_\text{sat}} = \frac{\rho_\ell(\mathrm{RH}=1)}{\rho_\text{sat}} \exp{\left[-\frac{1}{RT} \int_\mathrm{RH}^1 \frac{\partial \mu_{2,\ell}}{\partial \mathrm{RH}} \mathrm{d}\mathrm{RH} \right]}
\end{equation}
\noindent For large pores or nanopores connected to hierarchical larger pores, we can derive from Kelvin-Laplace equation that
\begin{equation}
\left(\frac{\partial \mu_{2,\ell}}{\partial P}\right) = \frac{v_2 \rho_\ell RT}{\mathrm{RH}}.
\end{equation}
\noindent In that case, we recover Krichevsky and Kasarnovsky equation expressed in terms of relative humidity:
\begin{equation}
\frac{c(\mathrm{RH})}{c_\text{sat}} = \frac{\rho_\ell(\mathrm{RH}=1)}{\rho_\text{sat}} \mathrm{e}^{-\rho_\ell v_2 \ln{\mathrm{RH}}}.
\end{equation}

\section*{5. More refined model}

To refine Krichevsky and Kasarnovsky relation, one can account for a variation of the partial molar volume $v_2$ with pressure by using more accurate equations of state for the liquid phase, as described by Widom and Cerdeiri{\~n}a~\cite{widom_2012,cerdeirina_2016}. For instance, we fit the density curve $\rho(P)$ provided by the IAPWS equation of state~\cite{wagner_2002,davitt_2010} for pressures between $0$ and $-\SI{80}{\mega\pascal}$ with van der Waals-like equation of states. This results in an agreement between the fit and the IAPWS data of a few percent with interaction parameter and covolumes respectively equal to $a_1=7.14 \times 10^{-49} \, \mathrm{Pa \cdot m^6}$ and $b_1=\SI{2.49}{\meter\cubed}$. It is to note that deviations from IAPWS in the negative pressure regime have been observed below $-\SI{50}{\mega\pascal}$~\cite{pallares_2016}.

Then, we assume that the (dilute) aqueous solution of dioxygen follows the same equation of state with effective parameters determined by an empirical mixing rule:~\cite{vankonynenburg_1980}
\begin{equation}
\left\{ \begin{array}{l}
a = [\rho_1^2 a_1 + 2 \rho_1 \rho_2 a_{12} + \rho_2^2 a_2] / (\rho_1+\rho_2)^2 \\
b = [\rho_1^2 b_1 + 2 \rho_1 \rho_2 b_{12} + \rho_2^2 b_2] / (\rho_1+\rho_2)^2
\end{array} \right. 
\end{equation}
\noindent where $\rho_1$ and $\rho_2 \, [\SI{}{\per\meter\cubed}]$ are the number densities of solvent and solute. Parameters $a_1$, $a_2$, $b_1$ and $b_2$ are respectively the interaction parameters ($a$) and covolumes ($b$) of pure water (subscript $1$) and pure oxygen (subscript $2$. The parameters $a_2$ and $b_2$ play no role in the considered dilute limit.  The parameters $a_{12}$ and $b_{12}$ are fitting parameters used to enforce agreement with tabulated data at saturation.

With these assumptions, the solubility parameter $\Sigma(P,T)$ introduced by Widom and defined as the ratio of $\rho_2$ over the value it would have in a hypothetical ideal gas at the same temperature and composition, can be computed. In the dilute limit, the concentration of dioxygen is finally expressed as
\begin{equation}
c = \frac{\Sigma P_\mathrm{O_2}}{RT}.
\end{equation}
\noindent Detailed calculation of $\Sigma$ for different Van der Waals-like equations of state have been presented by Widom and Cerdeiri{\~n}a ~\cite{cerdeirina_2016}. Parameters $a_{12}$ and $b_{12}$ are chosen to match the tabulated values of solubility~\cite{handbook} $\Sigma=3.1 \cdot 10^{-2}$ and oxygen molar volume~\cite{taylor_1978,zhou_2001} $v_2=\SI{3.21e-5}{\meter\cubed\per\mole}$ at saturation and standard pressure.

\begin{figure}[!htb]
\centering
\includegraphics{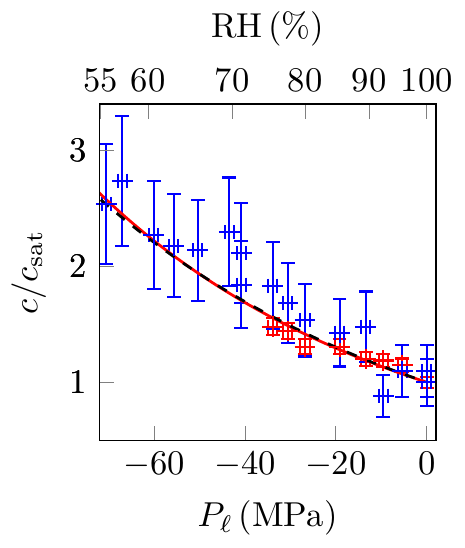}
\caption{Evolution of oxygen concentration $c$ with liquid pressure $P_\ell$ in metastable water, with respect to its value $c_\text{sat}$ at saturation. Red symbols correspond to measurements in Vycor glass and blue symbols to measurements in porous silicon. Black dashed line is the prediction of the simple model presented in main text while red continuous line is the prediction by the more precise model. \label{fig:conc_vs_RH}}
\end{figure}

Prediction of this model is compared to these of the simple model presented in the main text and to the experimental results. Within experimental uncertainty, we cannot distinguish between predictions by the two models. As well as for the agreement with the model by Lassin et al.~\cite{helgeson_1981,lassin_2016}, this can be explained by the fact that the molar volume of oxygen in water is almost independent on pressure in the considered range~\cite{mercury_2003}.

\providecommand{\latin}[1]{#1}
\makeatletter
\providecommand{\doi}
  {\begingroup\let\do\@makeother\dospecials
  \catcode`\{=1 \catcode`\}=2 \doi@aux}
\providecommand{\doi@aux}[1]{\endgroup\texttt{#1}}
\makeatother
\providecommand*\mcitethebibliography{\thebibliography}
\csname @ifundefined\endcsname{endmcitethebibliography}
  {\let\endmcitethebibliography\endthebibliography}{}

\end{document}